\documentclass{aa}  

\usepackage{graphicx}
\usepackage{txfonts}

\begin{document}

\title{Characterization of the GeV emission from the Kepler supernova remnant }

   \author{F. Acero\inst{1}
          \and
          M. Lemoine-Goumard\inst{2}
          \and
          J. Ballet\inst{1}
          }

   \institute{AIM, CEA, CNRS, Universit\'e Paris-Saclay, Universit\'e de Paris, F-91191 Gif sur Yvette, France \\
          \email{fabio.acero@cea.fr}
         \and
             Univ.  Bordeaux, CNRS, CENBG, UMR 5797, F-33170 Gradignan, France\\
             }

   \date{}

  \abstract{The Kepler supernova remnant (SNR) is the only historic supernova remnant lacking a detection at GeV and TeV energies which probe particle acceleration. A recent analysis of \emph{Fermi}-LAT data reported a likely GeV $\gamma$-ray candidate in the direction of the SNR.
Using approximately the same dataset but with an optimized analysis configuration, we confirm the $\gamma$-ray candidate to a solid $>6\sigma$ detection and report a spectral index of $2.14 \pm 0.12_{\rm stat} \pm 0.15_{\rm syst}$ for 
an energy flux above 100 MeV of $(3.1 \pm 0.6_{\rm stat} \pm 0.3_{\rm syst}) \times 10^{-12}$ erg~cm$^{-2}$~s$^{-1}$.
The $\gamma$-ray excess is not significantly extended and is fully compatible with the radio, infrared or X-ray spatial distribution of the SNR.
We successfully characterized this multi-wavelength emission with a model in which accelerated particles interact with the dense circumstellar material in the North-West portion of the SNR and radiate GeV $\gamma$-rays through $\pi^{o}$ decay.
The X-ray synchrotron and inverse-Compton (IC) emission mostly stem from the fast shocks in the southern regions with a  magnetic field B$\sim$100 $\mu$G or higher.
Depending on the exact magnetic field amplitude, the TeV emission could arise from either the South region (IC dominated) or the interaction region ($\pi^{o}$ decay dominated).}

   \keywords{  supernovae: individual : Kepler -- ISM: supernova remnants -- ISM: cosmic rays -- Gamma rays: general -- Astroparticle physics  --Shock waves   }

   \maketitle
%

\section{Introduction} \label{sec:intro}

The last Galactic supernova to be observed from Earth occurred on October 9, 1604 and a detailed report was produced by Johannes Kepler whose name is now attached to the supernova and its remnant. The \object{Kepler SNR} is most certainly the remnant of a Type Ia explosion but the large scale asymmetry with brighter emission towards the North from radio to X-rays \citep{2002ApJ...580..914D,2004A&A...414..545C,2007ApJ...662..998B,2007ApJ...668L.135R} has caused some confusion with a core collapse origin \citep[see][for a review]{2017hsn..book..139V}.
This asymmetry is now thought to be associated with circumstellar medium (CSM) from a runaway supernova progenitor system with significant mass loss prior to the explosion in a single degenerate scenario  \citep[e.g.][]{1987ApJ...319..885B,2013ApJ...764...63B,2015ApJ...808...49K}.

Estimates for the distance to the SNR range widely, from 3 to 7\, kpc in the literature \citep[e.g.][]{1999AJ....118..926R,2005AdSpR..35.1027S, 2008ApJ...689..225K}. The measurement of the proper motion of Balmer-dominated filaments  using the Hubble space telescope at a 10-year interval combined with the independently derived shock velocity from spectroscopy (H$\alpha$ line width) provides the most robust estimation at $d=5.1_{-0.7}^{+0.8}$ kpc \citep{2016ApJ...817...36S}.   
Throughout the paper we will use a distance of 5 kpc and rescale the values from the literature (e.g. shock speed) to match this distance whenever possible.

In the X-ray band, the emission is dominated by thermal emission with strong lines and in particular Fe lines supporting a Type Ia origin \citep[e.g.][]{2004A&A...414..545C, 2007ApJ...668L.135R}. 
Non-thermal emission from thin synchrotron-dominated filaments was later revealed by Chandra observations \citep{2005ApJ...621..793B,2007ApJ...668L.135R}.
Proper motion studies of these synchrotron rims \citep{2008ApJ...689..231V,2008ApJ...689..225K} show
fast shocks with velocities\footnote{Velocities were rescaled to a 5 kpc distance.} ranging from $\sim$2000 km s$^{-1}$ in the northern region to $\sim$5000 km s$^{-1}$ in the South.

The slower velocities in the North are related to the higher CSM density in this direction.
Measurement of the thickness of these filaments suggests a high magnetic field of 150-300 $\mu$G if the width is energy loss limited \citep{2005ApJ...621..793B, 2006A&A...453..387P}.

\begin{figure*}[ht!]

\includegraphics[width=0.31\textwidth]{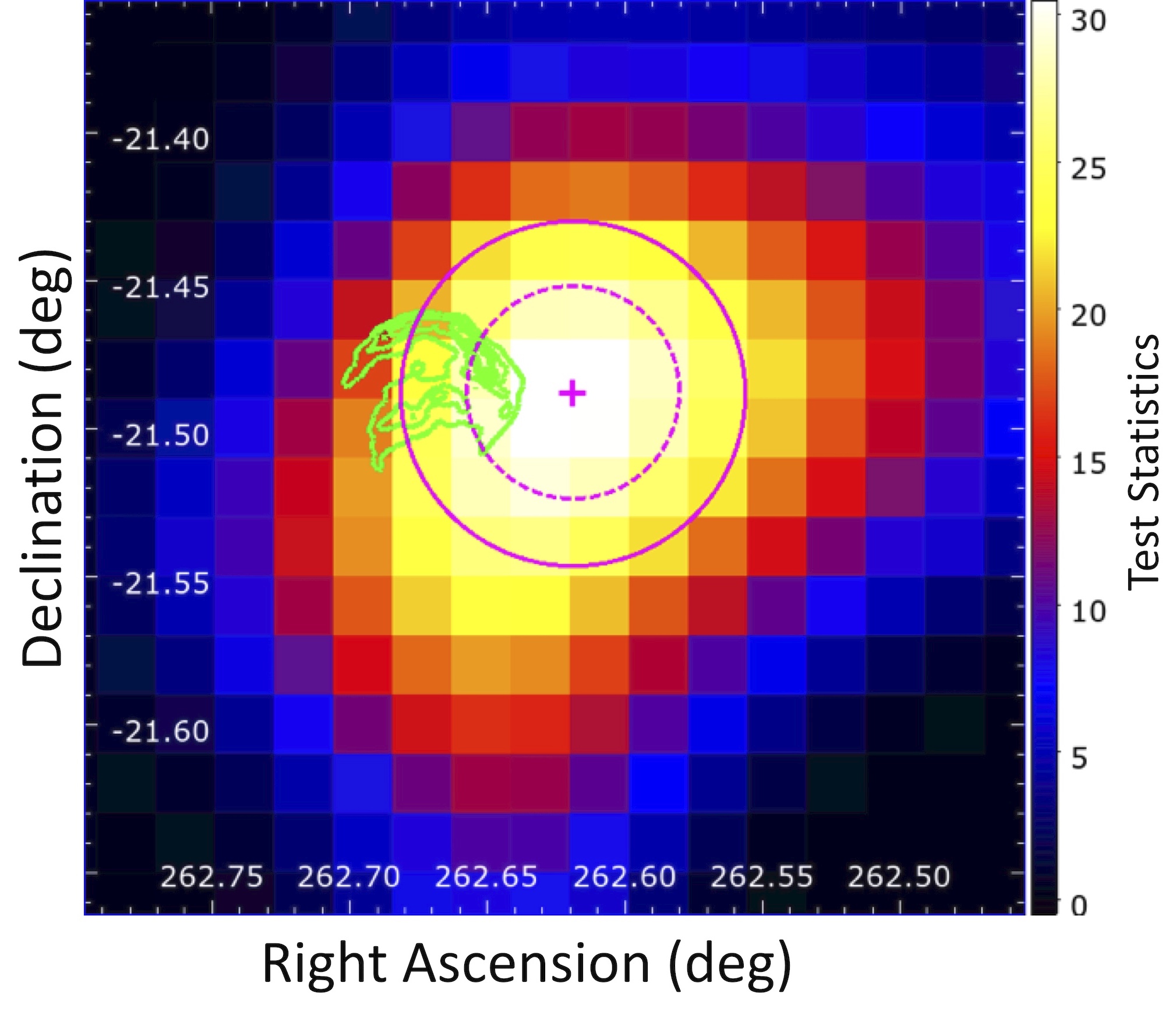}
\includegraphics[width=0.31\textwidth]{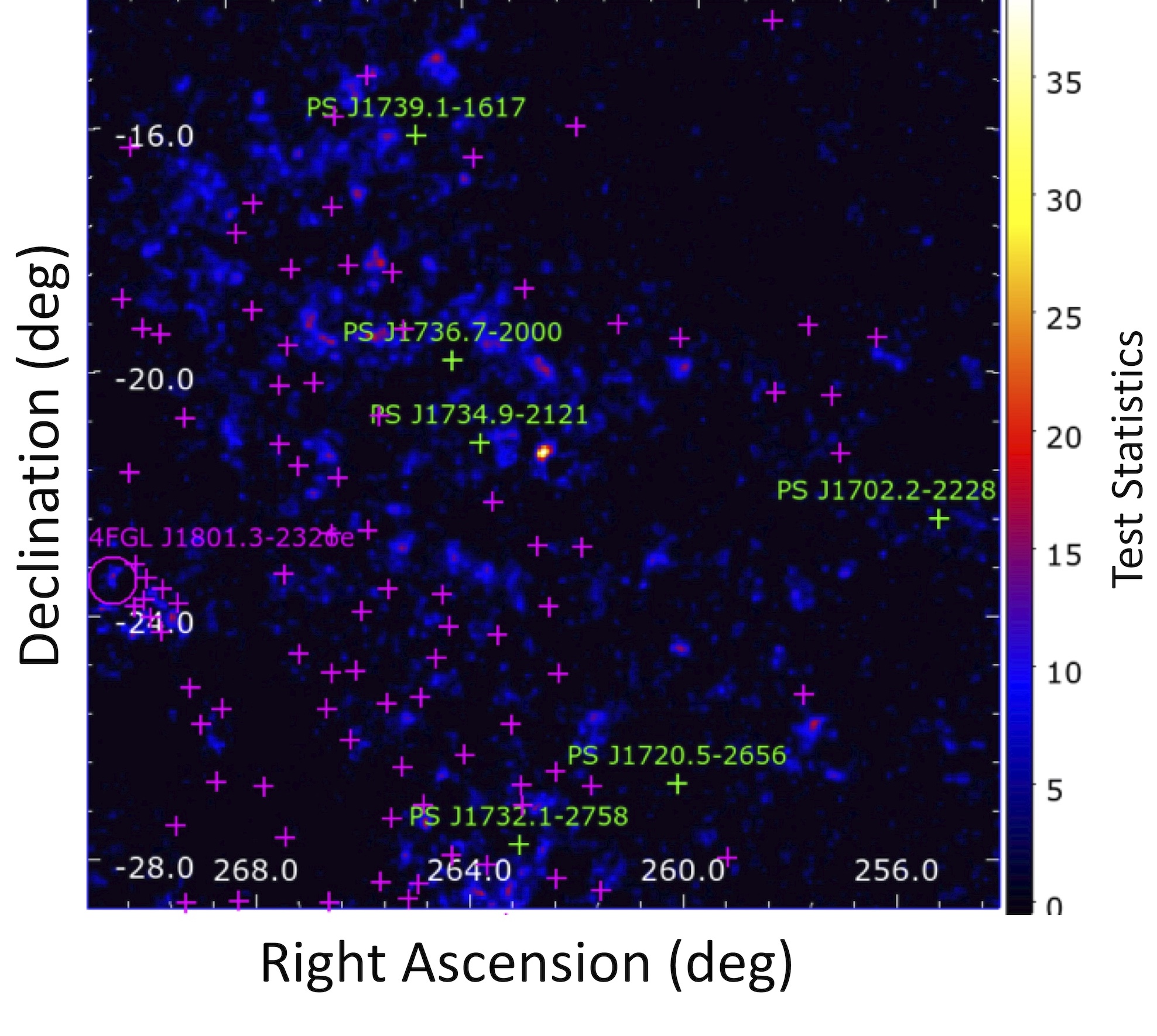}
\includegraphics[width=0.38\textwidth]{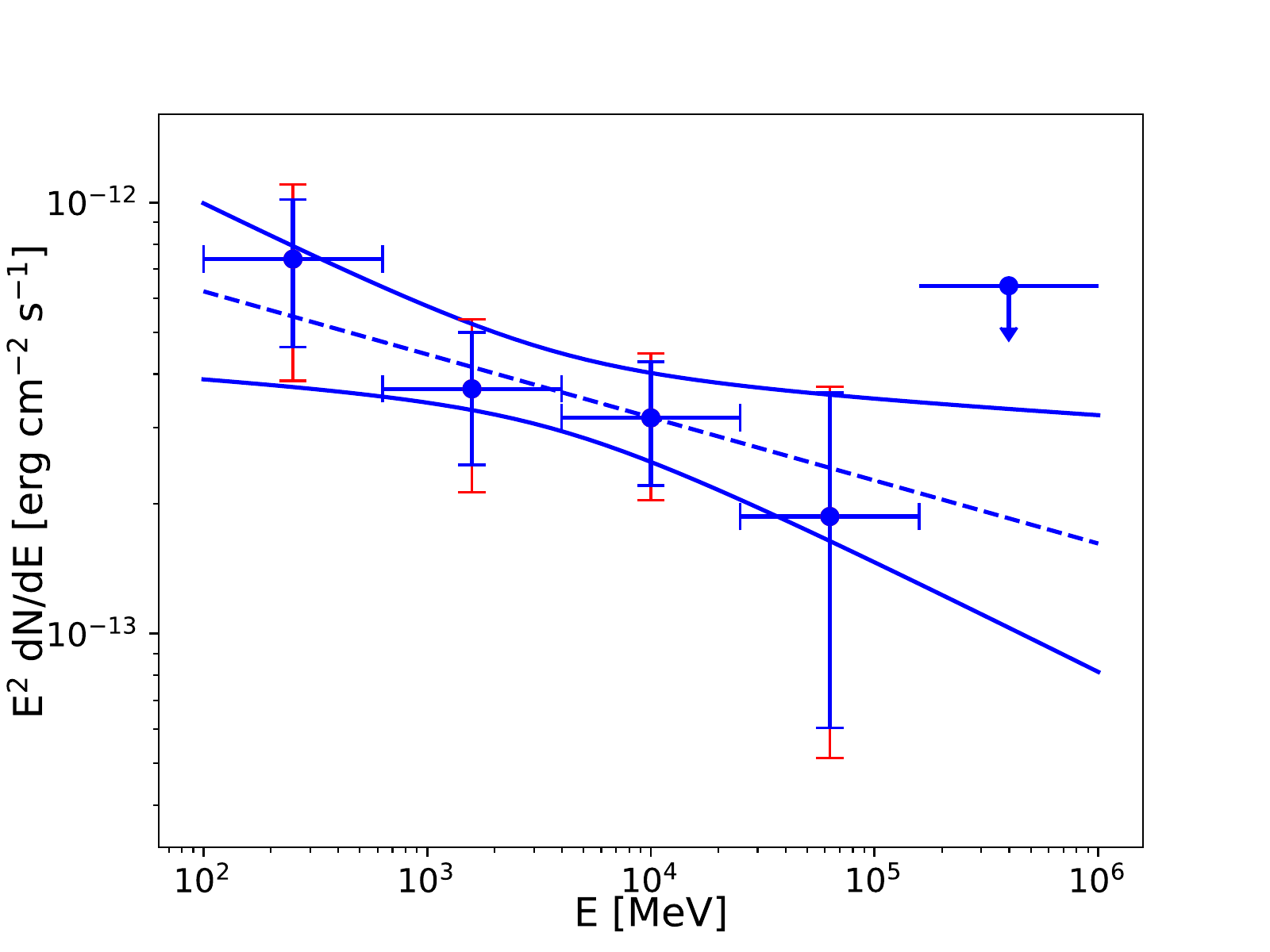}

\caption{Left panel: Zoom on the \emph{Fermi}-LAT TS map at the Kepler SNR position above 1 GeV. The green contours are from the infrared 24 $\mu$m \textit{Spitzer} map. The plus symbol and circles illustrate the best-fit position and the 68\%/95\% confidence contours. Middle panel: \emph{Fermi}-LAT TS map of the $15\degr \times 15\degr$ region of interest around the Kepler SNR above 100 MeV.  Magenta plus symbols represent the sources from the 4FGL-DR2 catalog and the green symbols  the added point sources. 
For both TS maps, the Kepler SNR was not included in the model.
Right panel: Spectral energy distribution of the Kepler SNR obtained using the infrared spatial template. Blue error bars represent the statistical uncertainties, while the  red ones correspond to the statistical and systematic uncertainties added in quadrature.}
\label{fig:tsmaps}
\end{figure*}

Despite being one of the youngest SNRs in our Galaxy with  high velocity shocks and signs of dense material and interaction, Kepler was the only historic SNR  without detected $\gamma$-ray emission until now.
This has changed with the recent report by \citet{2021ApJ...908...22X} of a $\sim 3.8 \, \sigma$ detection\footnote{Significance associated to a Test Statistic of 22.94 with 4 degrees of freedom.}  with \emph{Fermi}-LAT in the direction of the Kepler SNR.
In addition, the recent detection of TeV $\gamma$-rays after a deep exposure with the H.E.S.S. telescopes \citep[152 hours,][H.E.S.S. collaboration submitted]{2021icrc} opens the window for a  detailed $\gamma$-ray study of this young and historic SNR.

In this work we aim to transform the status of the \emph{Fermi}-LAT discovery from likely candidate to solid  detection by using a more sophisticated  analysis with approximately the same dataset (see Sect.~\ref{sect:analysis}). 
In addition to the modest significance, \citet{2021ApJ...908...22X} find a slightly offset best-fit position from the SNR.  We thus analyze in detail if this offset is statistically compatible with the SNR morphology, as realized by multi-wavelength spatial templates.
We conclude in Sect.~\ref{sect:discussion} by modeling Kepler's multi-wavelength emission under the assumption that $\gamma$-rays are emitted from the northern interacting region while synchrotron and inverse Compton emission arises mostly from the fast shocks in the southern region.

\section{Analysis}
\label{sect:analysis}
\subsection{LAT data reduction and preparation}
The \emph{Fermi}-LAT is a $\gamma$-ray telescope which detects photons by conversion into electron-positron pairs in the range from 20 MeV to higher than 500 GeV \citep{2009ApJ...697.1071A}. The following analysis was performed using 12 years of \emph{Fermi}-LAT data (2008 August 04 -- 2020 August 03). A maximum zenith angle of 90$^{\circ}$ below 1 GeV and 105$^{\circ}$ above 1 GeV was applied to reduce the contamination of the Earth limb, and the time intervals during which the satellite passed through the South Atlantic Anomaly were excluded. Our data were also filtered removing time intervals around solar flares and bright GRBs. The data reduction and exposure calculations were performed using the LAT $fermitools$ version 1.2.23 and $fermipy$ \citep{2017ICRC...35..824W} version 0.19.0. We performed a binned likelihood analysis and accounted for the effect of energy dispersion (when the reconstructed energy differs from the true energy) by using ${\rm edisp\_bins}=-3$. This means that the energy dispersion correction operates on the spectra with three extra bins below and above the threshold of the analysis\footnote{\url{https://fermi.gsfc.nasa.gov/ssc/data/analysis/documentation/Pass8_edisp_usage.html}}. Our binned analysis is performed with 10 energy bins per decade, spatial bins of $0.02^{\circ}$ for the morphological analysis and $0.05^{\circ}$ for the spectral analysis over a region of $15^{\circ} \times 15^{\circ}$. We included all sources from the LAT 10-year Source Catalog (4FGL-DR2\footnote{\url{https://fermi.gsfc.nasa.gov/ssc/data/access/lat/10yr_catalog}}) up to a distance of $15^{\circ}$ from Kepler.
Sources with a predicted number of counts below 1 and a significance to zero were removed from the model.

The summed likelihood method was used to simultaneously fit events with different angular reconstruction quality (PSF0 to PSF3 event types\footnote{\url{https://fermi.gsfc.nasa.gov/ssc/data/analysis/documentation/Cicerone/Cicerone_Data/LAT_DP.html}}). The Galactic diffuse emission was modeled by the standard file gll\_iem\_v07.fits and the residual background and extragalactic radiation were described by a single isotropic component with the spectral shape in the tabulated model iso\_P8R3\_SOURCE\_V3\_v1.txt. The models are available from the \emph{Fermi} Science Support Center (FSSC)\footnote{\url{https://fermi.gsfc.nasa.gov/ssc/data/access/lat/BackgroundModels.html}}.

Since the point spread function (PSF) of the \emph{Fermi}-LAT is energy dependent and broad at low energy, we started the morphological analysis at 1 GeV while the spectral analysis was made from 100 MeV up to 1 TeV. 

\subsection{Morphological analysis}\label{section:morpho}
The spectral parameters of the sources in the model were first fit simultaneously with the Galactic and isotropic diffuse emissions from 1 GeV to 1 TeV. During this procedure, a point source fixed at the position (RA$_{\rm J2000}$, Dec$_{\rm J2000}$ = $262.62^{\circ}, -21.49^{\circ}$) reported by \citet{2021ApJ...908...22X} was used to reproduce the $\gamma$-ray emission of the Kepler SNR. To search for additional sources in the region of interest (ROI), we computed a test statistic (TS) map that tests at each pixel the significance of a source with a generic E$^{-2}$ spectrum against the null hypothesis: ${\rm TS}=2(\ln \mathcal{L}_1 - \ln \mathcal{L}_0)$, where $\mathcal{L}_0$ and $\mathcal{L}_1$ are the likelihoods of the background (null hypothesis) and the hypothesis being tested (source plus background). We iteratively added four point sources in the model where the TS exceeded 25. We localized the four additional sources (RA$_{\rm J2000}$, Dec$_{\rm J2000}$ = $255.56^{\circ}, -22.47^{\circ}$; $260.13^{\circ}, -26.95^{\circ}$; $263.04^{\circ}, -27.97^{\circ}$,  and $264.78^{\circ}, -16.30^{\circ}$) and we fit their power-law spectral parameters. We then localized the source associated with Kepler and we obtained our best Point Source model (PS) at RA$_{\rm J2000}$ = 262.618$^{\circ}$ $\pm$ 0.023$^{\circ}$, Dec$_{\rm J2000}$ = -21.488$^{\circ}$ $\pm$ 0.026$^{\circ}$. The radii at 68\% confidence and 95\% confidence provided by $fermipy$ are: 0.036$^{\circ}$ and 0.058$^{\circ}$. During this fit, we left free the normalization of sources located closer than $4^{\circ}$ from the ROI center as well as the Galactic and isotropic diffuse emissions. Figure~\ref{fig:tsmaps} (left panel) presents our best localization and confidence radii superimposed on the TS map above 1 GeV. We tested its extension by localizing a 2D symmetric Gaussian. The significance of the extension is calculated through ${\rm TS_{\rm ext}}=2(\ln \mathcal{L}_{\rm ext} - \ln \mathcal{L}_{\rm PS})$ where $\mathcal{L}_{\rm ext}$ and $\mathcal{L}_{\rm PS}$ are the likelihood obtained with the extended and point-source model, respectively. For Kepler, we found ${\rm TS_{\rm ext}}=0.3$ indicating that the emission is not significantly extended compatible with the result from \citet{2021ApJ...908...22X}. The 95\% confidence level upper limit on the extension is 0.09$^{\circ}$ (for comparison the radius of the SNR is 0.03$^{\circ}$).

 Finally, we also examined the correlation of the $\gamma$-ray emission from Kepler with multi-wavelength templates derived from the VLA at 1.4 GHz \citep{2002ApJ...580..914D}, \textit{Spitzer} in infrared at 24 $\mu$m \citep{2007ApJ...662..998B}, and $Chandra$ in the 0.5-4 keV energy band \citep{2007ApJ...668L.135R}. We cannot use the likelihood ratio test to compare the hypothesis of a point source model to that of a multi-wavelength template, because the two models are not nested. However, we can use the Akaike information criterion \citep[AIC;][]{1974ITAC...19..716A} which takes into account the number k of independently adjusted parameters in a given model. The standard AIC formula is AIC = $2k - 2\rm{\ln \mathcal{L}}$. Here we estimate a $\Delta$AIC = $2k - \rm{TS}$ comparing the AIC of the null source hypothesis and the AIC of the source being tested.
    
The lowest $\Delta$AIC value, reported in Table~\ref{tab:morpho}, is obtained for the infrared template from \textit{Spitzer}.
  While a lower $\Delta$AIC indicates statistical preference for a model, the similar $\Delta$AIC values indicate that each of these models provides an equally good representation of the $\gamma$-ray signal detected by the LAT.  This result is consistent with our expectations from the \emph{Fermi}-LAT PSF ($\sim$ 1$\degr$ at 1 GeV\footnote{\url{https://www.slac.stanford.edu/exp/glast/groups/canda/lat_Performance.htm}}) and the 0.03$\degr$ SNR radius: all templates have a similar structure when convolved with the relatively broad PSF.  Hence, in our analysis below, we adopt the infrared template but emphasise that any of these templates would yield similar residual TS maps or inferred spectral properties.

\begin{table}
\caption{Results of the fit of the LAT data between 1 GeV and 1 TeV using different spatial models. The second column reports the TS values obtained for each spatial model, while column 3 indicates the number of degrees of freedom k adjusted in the model. The delta Akaike criterion is reported in the fourth column. See Sect.~\ref{section:morpho} for more details.}
\label{tab:morpho} 
\centering    
\begin{tabular}{lccc}
\hline \hline
Spatial model & TS & k & $\Delta$AIC \\
\hline
X-ray template        & 28.6    & 2 & -24.6  \\
Radio template       &   28.8  & 2 & -24.8 \\
Infrared template     & 29.7  & 2 & -25.7 \\
Best point source (PS)  &  32.1  & 4 & -24.1 \\
\end{tabular}
\end{table}

\begin{table}
\caption{Impact on the source significance of different analysis setups above a 700 MeV energy threshold. For comparison our setup corresponds to configuration 1 and that of \citet{2021ApJ...908...22X} to configuration 4.}
\label{tab:comp}
\centering    
\begin{tabular}{clccc}
\hline \hline
Configuration & Summed &  Bin size & Region size & TS \\
number & analysis & ($^{\circ}$) & ($^{\circ}$) & \\ 
\hline
1 & Yes & 0.05     & 15 &  33.9 \\
2 & No     & 0.05  & 15 & 30.6 \\
3 & No       & 0.1    & 15 & 23.2  \\
4 & No  &  0.1  & 20 & 21.4 \\
\end{tabular}
\end{table}

\subsection{Spectral analysis}\label{section:spec}
Using the best-fit infrared spatial template, we performed the spectral analysis from 100 MeV to 1 TeV.  We first verified whether any additional sources were needed in the model by examining the TS maps above 100 MeV. Two additional sources were detected at RA$_{\rm J2000}$, Dec$_{\rm J2000}$ = $264.19^{\circ}, -20.01^{\circ}$; $263.72^{\circ}, -21.36^{\circ}$ (not detected in the 1 GeV -- 1 TeV range used in Sect.~\ref{section:morpho}). The TS map obtained with all the sources considered in the model (see Fig.~\ref{fig:tsmaps}, middle panel) shows no significant residual emission, indicating that the ROI is adequately modeled. We then tested different spectral shapes for Kepler. During this procedure, the spectral parameters of sources located up to $4^{\circ}$ from the ROI center were left free during the fit, like those of the Galactic and isotropic diffuse emissions. We tested a simple power-law model, a logarithmic parabola, and a smooth broken power-law model. Again, the improvement between the power-law model and the two other models is tested using the likelihood ratio test. In our case, $\Delta TS$ is 2.4 for the logarithmic parabola model and 3.1 for the smooth broken power-law representation, indicating that no significant curvature is detected. Assuming a power-law representation, the best-fit model for the photon distribution yields a TS of 38.3 above 100 MeV, a spectral index of $2.14 \pm 0.12_{\rm stat} \pm 0.15_{\rm syst}$, and a normalization of $(2.71 \pm 0.57_{\rm stat} \pm 0.26_{\rm syst})$ $\times 10^{-14}$ MeV$^{-1}$~cm$^{-2}$~s$^{-1}$ at the pivot energy of 2947 MeV. This implies an energy flux above 100 MeV of $(3.1 \pm 0.6_{\rm stat} \pm 0.3_{\rm syst}) \times 10^{-12}$ erg~cm$^{-2}$~s$^{-1}$ and a corresponding $\gamma$-ray luminosity of $(0.93 \pm 0.18 \pm 0.09) \times 10^{34}$ erg~s$^{-1}$ at a distance of 5 kpc.\\ 
The systematic errors on the spectral analysis depend on our uncertainties on the Galactic diffuse emission model, on the effective area, and on the spatial shape of the source. The first is calculated using eight alternative diffuse emission models following the same procedure as in the first \emph{Fermi}-LAT supernova remnant catalog  \citep{2016ApJS..224....8A}  and the second is obtained by applying two scaling functions on the effective area. We also considered the impact on the spectral parameters when changing the spatial model from the infrared template to the best point source hypothesis. These three sources of systematic uncertainties were added in quadrature.\\
The \emph{Fermi}-LAT spectral points shown in Fig.~\ref{fig:tsmaps} (right panel)  were obtained by dividing the 100 MeV -- 1 TeV energy range into 5 logarithmically-spaced energy bins and  performing a  maximum likelihood spectral analysis to estimate the photon flux in each interval, assuming a power-law shape with fixed photon index $\Gamma$=2  for the  source. The normalizations of the diffuse Galactic and isotropic emission were left free in each energy bin as well as those of the sources within $4^{\circ}$. A 95\% C.L. upper limit is computed when the TS value is lower than $1$. Spectral data points are given in Table~\ref{tab:fluxpoints}.\\
We examined the reason for the significant improvement of the derived TS value (of 38.3) with respect to the TS value of 22.9 above 700 MeV reported by \citet{2021ApJ...908...22X}. To do so, we re-analyzed the source above the same threshold 700 MeV using a point source localized at the position reported by \citet{2021ApJ...908...22X}.
Their setup corresponds to configuration 4 in 
Table~\ref{tab:comp} and we find a TS value very similar to theirs (22.9).
We tested several analyses with different spatial bin sizes, region sizes and with/without summed likelihood and the improvement step by step is shown in Table~\ref{tab:comp}. The higher TS value that we find in our analysis is most likely due to the summed likelihood analysis and the finer spatial binning of 0.05$^{\circ}$ (configuration 1). These improvements together with our lower energy threshold of 100 MeV boost our detection to the TS value of 38.3.

\begin{table}
\caption{\emph{Fermi}-LAT flux data points using the infrared template. The flux parameter with an asterisk denotes an upper-limit. The first (second) flux errors represent statistical (systematic) errors respectively.}
\label{tab:fluxpoints}
\centering    
\begin{tabular}{lcc}
\hline \hline
Energy band & $E^{2}dN/dE$ & TS \\
GeV & $10^{-13}$ erg cm$^{-2}$ s$^{-1}$ &  \\
\hline
0.25 (0.10-0.63)        &   7.39 (2.77, -2.77) (2.19, -2.32)     & 7.22 \\
1.58 (0.63-3.98)        &   3.70 (1.23, -1.30) (0.97, -1.03)     & 9.72 \\
10.00 (3.98-25.1)      &   3.17 (0.96, -1.11) (0.59, -0.68)     & 16.85 \\
63.10 (25.1-158)    &   1.87 (1.27, -1.75) (0.48, -0.65)    & 4.44 \\
398.11 (158-1000) &  6.41*                              & 0.65 \\

\end{tabular}
\end{table}

\section{Discussion}
\label{sect:discussion}

We now model the $\gamma$-ray emission in a multi-wavelength context. As Kepler is a well studied SNR, we aim to build a coherent model by fixing as many parameters as possible from observations and theoretical grounds.

\subsection{Model motivation}\label{sect:motivation}

Our assumption is that on the one hand, the observed GeV $\gamma$-ray emission is mostly of hadronic nature ($\pi^{0}$ decay) being radiated from the North-West hemisphere where the shock is in interaction with the dense CSM as traced by infrared and optical maps. On the other hand, the leptonic components (synchrotron and inverse Compton) arise from high velocity regions mostly observed in the South with shock speed\footnote{Velocities were rescaled from a distance of 4 kpc to 5 kpc to be consistent with our distance assumption.} ranging from 4000-7000 km s$^{-1}$ \citep[regions 4-12, ][]{2008ApJ...689..225K}.
X-ray synchrotron emission requires high-speed shock regions, but radio emission can be produced by slower shocks. However for simplicity we model the electron population with a single radio to X-ray population. Consequently we expect our model will underpredict the radio data points.

We use the measured and inferred properties of the SNR to fix various parameters of our model. First we fix the fast shocks (leptonic components) at 5000 km s$^{-1}$ and the slower shocks in the interacting region at 1700 km s$^{-1}$ for a target density of 8 cm$^{-3}$ as reported in \citet{2016ApJ...817...36S}.
Secondly we derive the electron spectral distribution assuming that the electron maximal energy is limited by  synchrotron losses and that proton maximal energy is limited by the age of the remnant.

\subsection{Theoretical context}\label{sect:theory}

Following the prescription of \citet{2006A&A...453..387P}, the acceleration timescale  for the particles to reach an energy $E$  at the forward shock can be written as:
\begin{equation}
\tau_{\mathrm{acc}} \simeq (30.6\,\mathrm{yr}) \frac{3r^{2}}{16(r-1)}\times k_{0}(E)\times E_{\mathrm{TeV}}\, B_{100}^{-1}\,V_{\mathrm{sh},3}^{-2}
\label{eq:tacc}
\end{equation}
where $r$ is the compression ratio assumed at the shock,  $B_{100}$ the downstream magnetic field in units of 100 $\mu$G, and $V_{\mathrm{sh},3}$ the shock speed in units of 1000 km s$^{-1}$.
The deviation to Bohm diffusion is parametrized by $k_{0} \geq 1$ defined as $D(E)=k_{0} D_{\rm Bohm}(E)$ where $D_{\rm Bohm}$ is the Bohm diffusion coefficient. At a value of one, the acceleration at the shock is the most efficient and the maximal reachable energy decreases for higher values of $k_{0}$.

In the loss limited regime, the electron maximal energy can be obtained by equating the acceleration timescale $\tau_{\mathrm{acc}}$ to the synchrotron loss time at the shock giving:

\begin{equation}
E_{\mathrm{e,max}} \simeq (8.3\,\mathrm{TeV})\times \bar{f}(r)\times k_{0}^{-1/2}\times B_{100}^{-1/2}\times V_{\mathrm{sh},3}
\label{eq:EeMax}
\end{equation}

where $\bar{f}(r) \equiv f(r)/f(4)$, with $f(r) = \sqrt{r-1}/r$.

The maximal proton energy is obtained by equating $\tau_{\mathrm{acc}}$ from Equation~\ref{eq:tacc} with the age of the remnant of 400 years giving:
\begin{equation}
E_{\mathrm{p,max}} \simeq (13.1\,\mathrm{TeV})\times T_{\rm 400} \times \bar{f}^{2}(r)\times k_{0}^{-1}\times B_{100}\, V^{2}_{\mathrm{sh},3}
\label{eq:EpMax}
\end{equation}
where $T_{\rm 400}$ is the age of the remnant in units of 400 yrs.
For the magnetic fields considered in this modeling ($B \sim$ 100 $\mu$G), the synchrotron cooling is non negligible and is modeled with a broken power-law with $E_{\rm break}$ obtained by equating the age of the remnant and the
synchrotron loss time downstream of the shock giving:
\begin{equation}
E_{\mathrm{break}} \simeq (3.1\,\mathrm{TeV})\times T_{\rm 400} \times  B^{-2}_{100}.
\label{eq:Eb}
\end{equation}

Because of the cooling and the cut at the maximum energy, the electron population is modeled as an \textit{Exponentially Cutoff BrokenPowerLaw} with a change of slope after $E_{\mathrm{break}}$ to $\Gamma_{2}=\Gamma_{1}+1$.

Assuming that the synchrotron emission is limited by cooling, the acceleration efficiency parameter $k_{0}$ can be indirectly estimated by comparing the shock speed and the cutoff energy of the X-ray synchrotron spectrum using Equation 34 of \citet{2007A&A...465..695Z}. 
Such a study was carried out by \citet{2021ApJ...907..117T} on a population of SNRs including Kepler.
For Kepler's south-eastern regions where fast shocks are observed \citep[][reg 4-8 in Table 2]{2021ApJ...907..117T}, $k_{0}$ (their $\eta$) ranges from 2.0 to 3.2 which is equivalent to 3.1 to 5.0\footnote{Velocity is derived from proper motion which depends linearly on the distance and $\eta$ has a square dependence on shock speed \citep[see Equation 3 from][]{2021ApJ...907..117T}.} when rescaled to a distance of 5 kpc instead of 4 kpc. 
Given the measured values mentioned above, we decided to fix $k_{0}=3.4$ (the median value of the distribution) for our modeling for both the electron and proton populations (radiative model shown in Fig.~\ref{fig:sed}).

The radiative models from the \textit{naima} packages \citep{2015ICRC...34..922Z} have been used with the \textit{Pythia8} parametrization of  \citet{2014PhRvD..90l3014K} for the $\pi^{o}$ decay.
For the IC, a far infrared field (T=30 K, $U_{\rm ph}$= 1 eV cm$^{-3}$) was used in addition to the CMB \citep{2006ApJ...648L..29P}.

\begin{table*}

\caption{List of parameters obtained from the modeling of the spectral energy distribution in different scenarios. Parameters in brackets are fixed from observables or theory (see Sect.~\ref{sect:motivation} and Sect.~\ref{sect:theory}) while other parameters are adjusted to the data. The energy budget values are integrated above 1 GeV.}

\begin{tabular}{lccccccccccc}
\hline \hline
Scenario & $B$ & $n_0$ & $V_{\rm sh,e}$ & V$_{\rm sh,p}$ & $\Gamma_{\rm e,1}/\Gamma_{\rm e,2}$ & $E_{\rm break,e}$ & $E_{\rm max,e}$ & $\Gamma_p$ & $E_{\rm max,p}$ & $W_{e}$ & $W_{p}$ \\
 & $\mu$G & cm$^{-3}$ & km s$^{-1}$ & km s$^{-1}$ &  & TeV & TeV &  & TeV & erg & erg \\
 \hline
High magnetic field  & 170 & [8] & [5000] & [1700] & 2.2/[3.2] & [1.1] & [18.4] & 2.2 & [21.2] & 1.7$\times 10^{47}$ & 5.6$\times 10^{48}$  \\
Intermediate magnetic field  & 90 & [8] & [5000] & [1700] & 2.3/[3.3] & [3.9] & [25.3] & 2.3 & [11.2] & 5.6$\times 10^{47}$ & 5.6$\times 10^{48}$ \\
\end{tabular}

\label{tab:param}
\end{table*}

\subsection{Multi-wavelength data and spectral energy distribution}

\begin{figure}[t!]

\includegraphics[width=1\columnwidth,bb=0 10 535 385,clip]{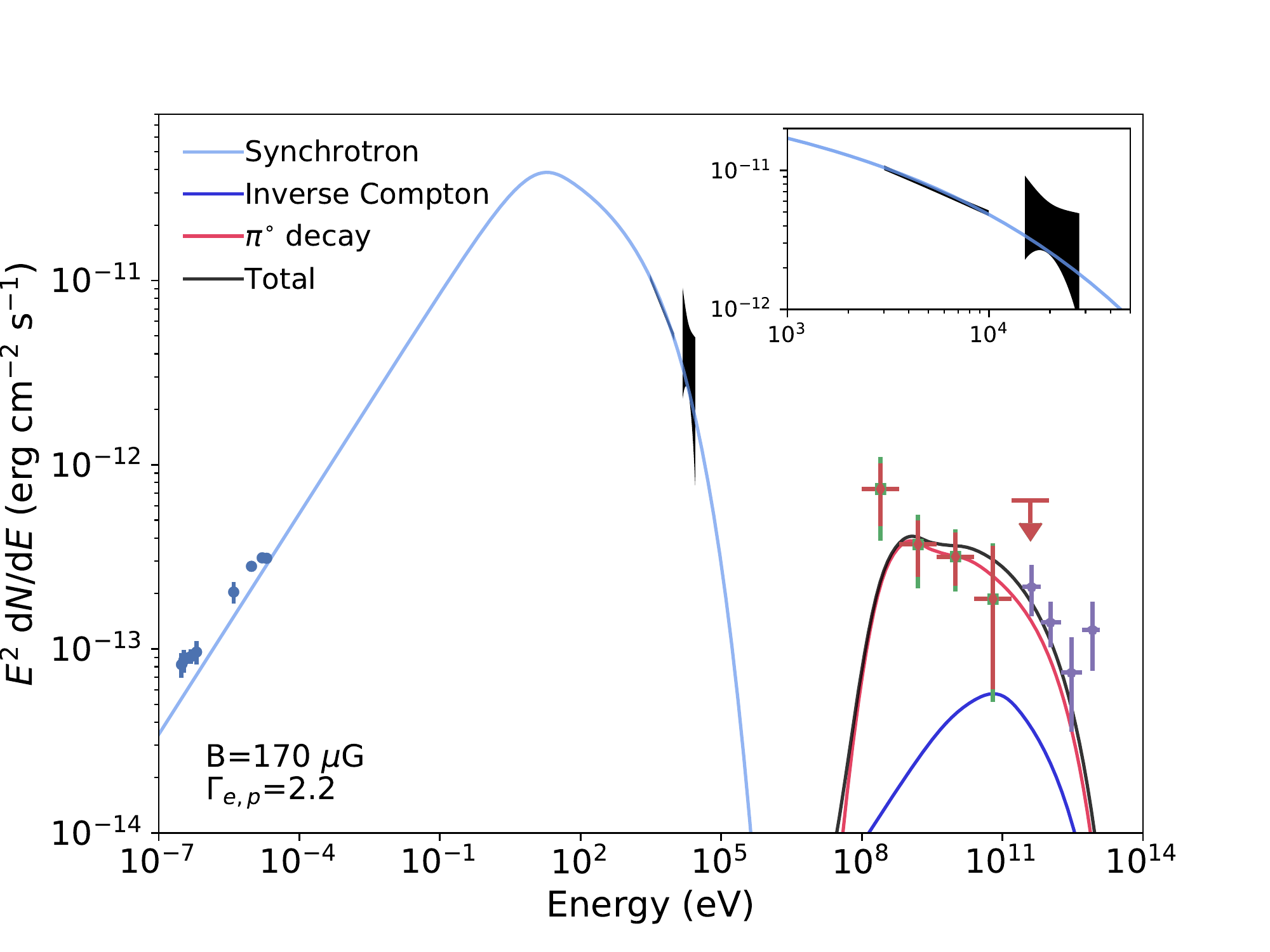}
\includegraphics[width=1\columnwidth,bb=0 10 535 385,clip]{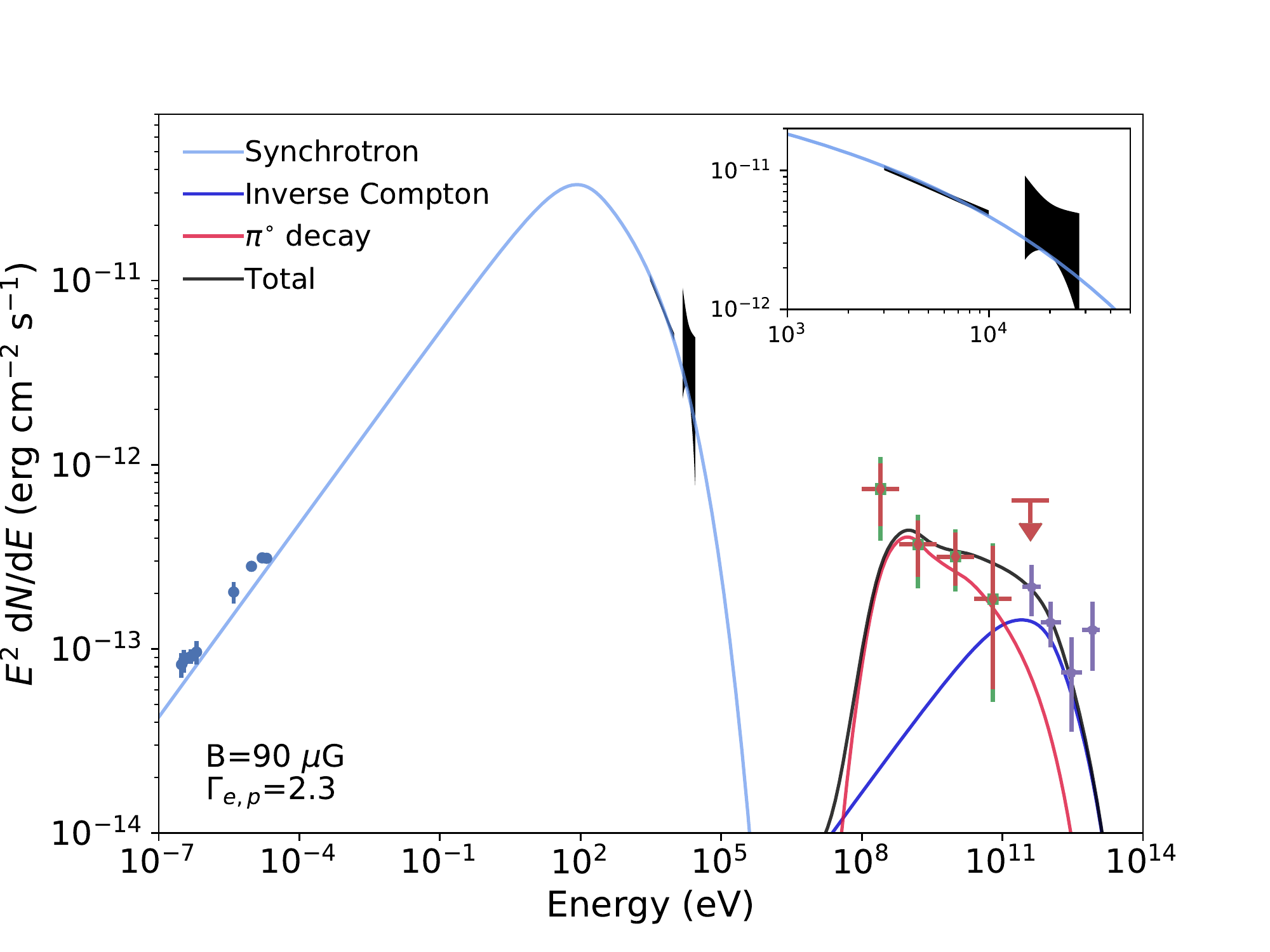}

\caption{Spectral energy modelling in an intermediate and high magnetic field scenario. The leptonic emission is assumed to arise from the fast shocks in the southern regions and the hadronic emission from the North-West interaction region. For the \emph{Fermi}-LAT flux points, both the statistical errors and summed errors ($\sqrt{stat^{2} +syst^{2}}$) are shown. Modeling parameters are listed in Table \ref{tab:param}. }
\label{fig:sed}
\end{figure}

For the multi-wavelength data presented in Fig.~\ref{fig:sed}, we used the updated compilation of radio fluxes from \citet{2021A&A...653A..62C}, the X-ray data from the \textit{Suzaku} XIS + HXD instruments \citep[covering the 3-10 keV and 15-30 keV band,][]{2021PASJ..tmp...10N}, and the H.E.S.S. flux points from \citet[][H.E.S.S. collaboration submitted]{2021icrc}.
The newly derived \emph{Fermi}-LAT flux points using the infrared spatial template are presented (same as Fig.~\ref{fig:tsmaps}).

The resulting adjusted models, shown in Fig.~\ref{fig:sed}, are obtained 
with only four free parameters being the downstream magnetic field, a unique electron/proton spectral index, and the associated energy budgets (see Table \ref{tab:param}).
The electron population spectral index and the amplitude of the magnetic field are correlated in the spectral energy distribution (SED) fitting.
 Motivated by theoretical expectations from recent kinetic hybrid simulations \citep{2021ApJ...922....1D} predicting spectral indices steeper than 2, we present two different scenarios for  spectral indices of 2.3 and 2.2 corresponding to an intermediate (90 $\mu$G) and high magnetic field value (170 $\mu$G) , respectively.
A viable scenario, not shown in Fig.~\ref{fig:sed}, can also be obtained for B=250 $\mu$G if the spectral index is changed to 2.15.
Such magnetic field value is compatible with the estimated values at the shock given the thin X-ray synchrotron filaments size \citep{2005ApJ...621..793B, 2012A&A...545A..47R}. 
When changing $k_{0}$ to 1 for the protons (i.e. maximal acceleration efficiency), $E_{\mathrm{p,max}}$ increases and slightly improves the model agreement with the H.E.S.S. flux points in the hadronic dominated case.

Assuming that the hadronic emission arises from a small angular region in the SNR could explain the modest energy budget required (5.6$\times 10^{48}$ erg). On the \textit{Spitzer} infrared 24 $\mu$m and the Hubble H$\alpha$ images from \citet{2016ApJ...817...36S}, the North-West interacting region has an opening angle of $\sim$45$\degr$. Assuming a similar angle in the third dimension, this spherical cap represents $\sim$15\% of the SNR surface. The local proton energy budget is therefore equivalent to about 4\% of the local kinetic energy assuming an energy explosion of $10^{51}$ erg.

The intermediate and high magnetic field scenarios reproduce equally well the GeV to TeV flux points and cannot be disentangled from the SED analysis alone.
However, we note that the inverse Compton emission dominates above 300 GeV in an intermediate magnetic field case while the hadronic emission dominates the entire $\gamma$-ray band for a high magnetic field scenario.
Therefore if the IC emission arises from the fast moving shocks in the southern regions, the precise location of the TeV $\gamma$-ray emission might be able to constrain the hadronic or leptonic nature of the emission and indirectly  the average magnetic field in the SNR.
The distance between the dense interacting region in the North-West and the southern rim is of the order of 0.05$\degr$. While this is at the limit of the H.E.S.S. telescopes source localization precision for a faint source, a comparison of the GeV and TeV best-fit positions could shed light on the nature of TeV $\gamma$-ray emission.
With an increased sensitivity and spatial resolution, the next generation Cherenkov Telescope Array \citep{2019scta.book.....C} will locate with great accuracy the Kepler SNR $\gamma$-ray emission.

\subsection{Kepler SNR $\gamma$-ray emission in context}

The detection of the Kepler SNR at GeV and TeV energies, completes our high-energy view of historical SNRs. 
In this section we compare the global spectral properties of the young and likely hadronic-dominated SNRs Kepler, Tycho, and Cassiopeia A with respect to the older middle-aged SNRs W 44, IC 443, and Cygnus Loop. Note that other young SNRs such as SN 1006, RX J1713.7$-$3946 or RCW 86, showing a spectral slope  $\Gamma \sim$ 1.5 at GeV energies \citep[see e.g.][]{2015A&A...580A..74A}, are likely dominated by leptonic emission and are not considered in our sample. The distances to the sources are fixed to 3.33 $\pm$ 0.10 kpc for Cassiopeia A \citep{2014MNRAS.441.2996A}, 4 $\pm$ 1 kpc for Tycho \citep{2010ApJ...725..894H}, 5.1$^{+0.8}_{-0.7}$ kpc for Kepler \citep{2016ApJ...817...36S},  735 $\pm$ 25 pc for Cygnus Loop \citep{2018MNRAS.481.1786F}, 3.0 $\pm$ 0.3 kpc for W 44 \citep{2018AJ....155..204R}, and 1.7 $\pm$ 0.1 kpc for IC 443 \citep{2019MNRAS.488.3129Y}.

\begin{figure}[t!]
\includegraphics[width=1\columnwidth,bb=5 10 535 385,clip]{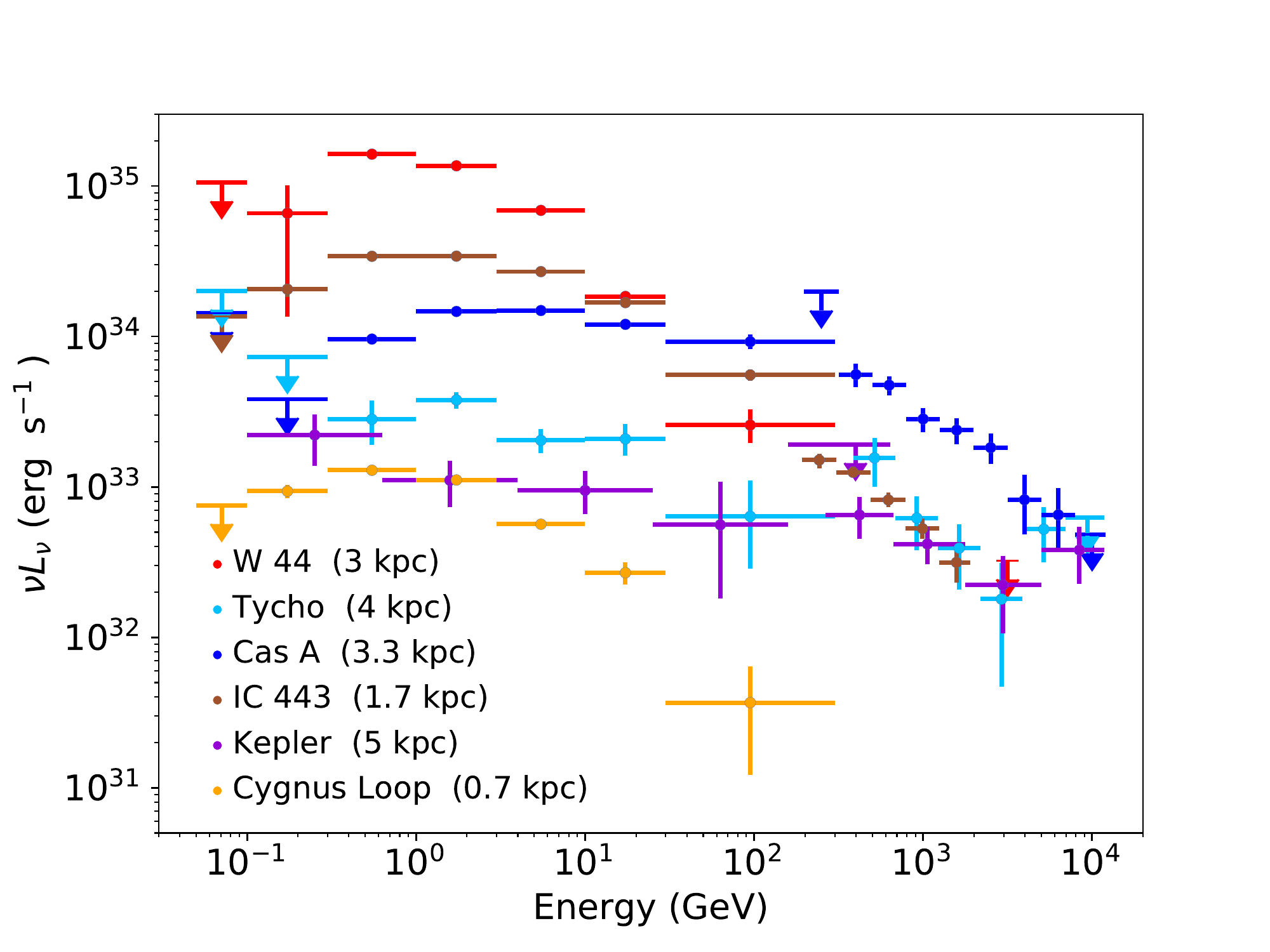}

\caption{Luminosity spectral energy distributions of a selection of SNRs for which the distance is well constrained and the $\gamma$-ray emission is likely dominated by hadronic emission. The  \emph{Fermi}-LAT data points are taken from the 4FGL DR2  catalog \citep{2020ApJ...892..105A} except for Kepler where data  from this work are used. TeV spectral data points for Cassiopeia A 
 are taken from \citet{2020ApJ...894...51A}, from \citet{2017ApJ...836...23A} for Tycho, from \citet[][ H.E.S.S. collaboration submitted]{icrc2021} for Kepler, from \citet{2015ICRC...34..875H} for IC 443, and from \citet{2018A&A...612A...1H} for W44 (upper-limit from the H.E.S.S. galactic plane survey). The references for the distances are given in the main text.}
\label{fig:luminosities}
\end{figure}

Figure~\ref{fig:luminosities} compares the SED in terms of luminosity of the aforementioned SNRs. 
As a proxy to discuss spectral curvature, we estimated the hardness ratio HR=$\nu F_{\rm 1 \,TeV}$/$\nu F_{\rm 1 \, GeV}$. Tycho, Kepler, and  Cassiopeia A  exhibit a nearly flat spectrum (HR=0.2-0.4) while the curvature is stronger for IC 443 (HR=0.015) and  W 44 ($<\,2 \times 10^{-3}$).
Such a contrast is due to differences in the acceleration and emission mechanisms. In the young SNRs sample, high shock speeds (3000-6000 km s$^{-1}$) are observed producing highly energetic CRs interacting with circumstellar material. The second sample exhibits lower shock speeds (few 100 km s$^{-1}$) with the presence of radiative shocks where compression and re-acceleration of pre-existing CRs takes place producing a spectral break at lower energies than for the young SNR sample.
We note that the separation in terms of luminosity in our sample  is not as clear. While W 44 is 50-100 times more luminous at 1 GeV than Kepler and Tycho, it is also 100 times more luminous than Cygnus Loop at 1 GeV. This is related to the fact that the shock in W 44 is interacting with dense molecular environments \citep[few 100 cm$^{-3}$,][]{2013ApJ...768..179Y} whereas it is closer to $\sim$1-10 cm$^{-3}$ in Cygnus Loop \citep{2018MNRAS.481.1786F}.

\section{Conclusion}

By using $\sim$12 years of \emph{Fermi}-LAT data and a summed 
likelihood analysis with the PSF event types, we were able to confirm GeV $\gamma$-ray emission at a $>$6$\sigma$ detection level that is spatially compatible with the Kepler SNR. From the analysis of this $\gamma$-ray emission we draw the following conclusions:

\begin{itemize}
\setlength{\itemsep}{0\baselineskip}

    \item[$\bullet$] above 100 MeV, the source is detected with a TS=38.3 with a power-law index of $2.14 \pm 0.12_{\rm stat} \pm 0.15_{\rm syst}$.
    
    \item[$\bullet$] the source is not significantly extended with an upper  limit  on  its  extension  of  0.09$^{\circ}$ (SNR radius is 0.03$^{\circ}$).
    
    \item[$\bullet$] the SED is modeled in a scenario with only four free parameters (B, $\Gamma_{\rm e,p}$, $W_{\rm e}$, $W_{\rm p}$), the rest being fixed from the literature and theoretical grounds. The GeV $\gamma$-ray emission is interpreted as $\pi^{o}$ decay from the North-West interaction region. The TeV emission could be IC dominated ($B < 100 \, \mu G$; expected peak location in the South) or $\pi^{o}$ decay dominated ($B > 100 \, \mu G$; expected peak location in the North-West). While this is at the limit of current generation instruments, a comparison of the \emph{Fermi}-LAT and HESS best-fit positions and errors could help to disentangle the two scenarios.
    
    \item[$\bullet$] assuming a particle density of 8 cm$^{-3}$, derived from infrared observations, and that the interaction region represents 15$\%$ of the SNR surface, the local fraction of kinetic energy transferred to accelerated particles is of the order 4$\%$.
    
\end{itemize}

\textit{Acknowledgments}: The authors would like to thank the \emph{Fermi}-LAT internal referee Francesco de Palma and the journal referee for comments and suggestions that helped improved the clarity of the paper. FA would like to acknowledge the hospitality of Villa Port Magaud where part of this work was carried out. M.L.G. acknowledges support from Agence Nationale de la Recherche (grant ANR- 17-CE31-0014). The \emph{Fermi}-LAT Collaboration acknowledges generous ongoing support from a number of agencies and institutes that have supported both the development and the operation of the LAT as well as scientific data analysis. These include the National Aeronautics and Space Administration and the Department of Energy in the United States, the Commissariat \'a l'Energie Atomique and the Centre National de la Recherche Scientifique / Institut National de Physique Nucl\'eaire et de Physique des Particules in France, the Agenzia Spaziale Italiana and the Istituto Nazionale di Fisica Nucleare in Italy, the Ministry of Education, Culture, Sports, Science and Technology (MEXT), High Energy Accelerator Research Organization (KEK) and Japan Aerospace Exploration Agency (JAXA) in Japan, and the K. A. Wallenberg Foundation, the Swedish Research Council and the Swedish National Space Board in Sweden. Additional support for science analysis during the operations phase is gratefully acknowledged from the Istituto Nazionale di Astrofisica in Italy and the Centre National d'\'etudes Spatiales in France.

\textit{Softwares}: 
This research made use of \textit{astropy}, a community-developed  Python package for Astronomy \citep{astropy:2013, astropy:2018}, of \textit{fermipy} a Python package for the \emph{Fermi}-LAT analysis \citep{2017ICRC...35..824W},  and of \textit{gammapy}, a community-developed Python package for TeV gamma-ray astronomy \citep{gammapy:2017, gammapy:2019}. The \textit{naima} package was used for the modeling \citep{2015ICRC...34..922Z}.


\begin{thebibliography}{47}
\expandafter\ifx\csname natexlab\endcsname\relax\def\natexlab#1{#1}\fi

\bibitem[{{Abeysekara} {et~al.}(2020){Abeysekara}, {Archer}, {Benbow}, {Bird},
  {Brose}, {Buchovecky}, {Buckley}, {Chromey}, {Cui}, {Daniel}, {Das},
  {Dwarkadas}, {Falcone}, {Feng}, {Finley}, {Fortson}, {Gent}, {Gillanders},
  {Giuri}, {Gueta}, {Hanna}, {Hassan}, {Hervet}, {Holder}, {Hughes},
  {Humensky}, {Kaaret}, {Kar}, {Kelley-Hoskins}, {Kertzman}, {Kieda}, {Krause},
  {Krennrich}, {Kumar}, {Lang}, {Maier}, {Moriarty}, {Mukherjee},
  {Nievas-Rosillo}, {O'Brien}, {Ong}, {Park}, {Petrashyk}, {Pfrang}, {Pohl},
  {Pueschel}, {Quinn}, {Ragan}, {Reynolds}, {Richards}, {Roache}, {Sadeh},
  {Santander}, {Sembroski}, {Shahinyan}, {Sushch}, {Weinstein}, {Wilcox},
  {Wilhelm}, {Williams}, {Williamson}, {Zitzer}, \&
  {Ghiotto}}]{2020ApJ...894...51A}
{Abeysekara}, A.~U., {Archer}, A., {Benbow}, W., {et~al.} 2020, \apj, 894, 51

\bibitem[{{Acero} {et~al.}(2016){Acero}, {Ackermann}, {Ajello}, {Baldini},
  {Ballet}, {Barbiellini}, {Bastieri}, {Bellazzini}, {Bissaldi}, {Blandford},
  {Bloom}, {Bonino}, {Bottacini}, {Brandt}, {Bregeon}, {Bruel}, {Buehler},
  {Buson}, {Caliandro}, {Cameron}, {Caputo}, {Caragiulo}, {Caraveo},
  {Casandjian}, {Cavazzuti}, {Cecchi}, {Chekhtman}, {Chiang}, {Chiaro},
  {Ciprini}, {Claus}, {Cohen}, {Cohen-Tanugi}, {Cominsky}, {Condon}, {Conrad},
  {Cutini}, {D'Ammando}, {de Angelis}, {de Palma}, {Desiante}, {Digel}, {Di
  Venere}, {Drell}, {Drlica-Wagner}, {Favuzzi}, {Ferrara}, {Franckowiak},
  {Fukazawa}, {Funk}, {Fusco}, {Gargano}, {Gasparrini}, {Giglietto}, {Giommi},
  {Giordano}, {Giroletti}, {Glanzman}, {Godfrey}, {Gomez-Vargas}, {Grenier},
  {Grondin}, {Guillemot}, {Guiriec}, {Gustafsson}, {Hadasch}, {Harding},
  {Hayashida}, {Hays}, {Hewitt}, {Hill}, {Horan}, {Hou}, {Iafrate}, {Jogler},
  {J{\'o}hannesson}, {Johnson}, {Kamae}, {Katagiri}, {Kataoka}, {Katsuta},
  {Kerr}, {Kn{\"o}dlseder}, {Kocevski}, {Kuss}, {Laffon}, {Lande}, {Larsson},
  {Latronico}, {Lemoine-Goumard}, {Li}, {Li}, {Longo}, {Loparco}, {Lovellette},
  {Lubrano}, {Magill}, {Maldera}, {Marelli}, {Mayer}, {Mazziotta}, {Michelson},
  {Mitthumsiri}, {Mizuno}, {Moiseev}, {Monzani}, {Moretti}, {Morselli},
  {Moskalenko}, {Murgia}, {Nemmen}, {Nuss}, {Ohsugi}, {Omodei}, {Orienti},
  {Orlando}, {Ormes}, {Paneque}, {Perkins}, {Pesce-Rollins}, {Petrosian},
  {Piron}, {Pivato}, {Porter}, {Rain{\`o}}, {Rando}, {Razzano}, {Razzaque},
  {Reimer}, {Reimer}, {Renaud}, {Reposeur}, {Rousseau}, {Saz Parkinson},
  {Schmid}, {Schulz}, {Sgr{\`o}}, {Siskind}, {Spada}, {Spandre}, {Spinelli},
  {Strong}, {Suson}, {Tajima}, {Takahashi}, {Tanaka}, {Thayer}, {Thompson},
  {Tibaldo}, {Tibolla}, {Torres}, {Tosti}, {Troja}, {Uchiyama}, {Vianello},
  {Wells}, {Wood}, {Wood}, {Yassine}, {den Hartog}, \&
  {Zimmer}}]{2016ApJS..224....8A}
{Acero}, F., {Ackermann}, M., {Ajello}, M., {et~al.} 2016, \apjs, 224, 8

\bibitem[{{Acero} {et~al.}(2015){Acero}, {Lemoine-Goumard}, {Renaud}, {Ballet},
  {Hewitt}, {Rousseau}, \& {Tanaka}}]{2015A&A...580A..74A}
{Acero}, F., {Lemoine-Goumard}, M., {Renaud}, M., {et~al.} 2015, \aap, 580, A74

\bibitem[{{Ajello} {et~al.}(2020){Ajello}, {Angioni}, {Axelsson}, {Ballet},
  {Barbiellini}, {Bastieri}, {Becerra Gonzalez}, {Bellazzini}, {Bissaldi},
  {Bloom}, {Bonino}, {Bottacini}, {Bruel}, {Buson}, {Cafardo}, {Cameron},
  {Cavazzuti}, {Chen}, {Cheung}, {Ciprini}, {Costantin}, {Cutini}, {D'Ammando},
  {de la Torre Luque}, {de Menezes}, {de Palma}, {Desai}, {Di Lalla}, {Di
  Venere}, {Dom{\'\i}nguez}, {Dirirsa}, {Ferrara}, {Finke}, {Franckowiak},
  {Fukazawa}, {Funk}, {Fusco}, {Gargano}, {Garrappa}, {Gasparrini},
  {Giglietto}, {Giordano}, {Giroletti}, {Green}, {Grenier}, {Guiriec},
  {Harita}, {Hays}, {Horan}, {Itoh}, {J{\'o}hannesson}, {Kovac'evic'},
  {Krauss}, {Kreter}, {Kuss}, {Larsson}, {Leto}, {Li}, {Liodakis}, {Longo},
  {Loparco}, {Lott}, {Lovellette}, {Lubrano}, {Madejski}, {Maldera},
  {Manfreda}, {Mart{\'\i}-Devesa}, {Massaro}, {Mazziotta}, {Mereu}, {Meyer},
  {Migliori}, {Mirabal}, {Mizuno}, {Monzani}, {Morselli}, {Moskalenko},
  {Negro}, {Nemmen}, {Nuss}, {Ojha}, {Ojha}, {Omodei}, {Orienti}, {Orlando},
  {Ormes}, {Paliya}, {Pei}, {Pe{\~n}a-Herazo}, {Persic}, {Pesce-Rollins},
  {Petrov}, {Piron}, {Poon}, {Principe}, {Rain{\`o}}, {Rando}, {Rani},
  {Razzano}, {Razzaque}, {Reimer}, {Reimer}, {Schinzel}, {Serini}, {Sgr{\`o}},
  {Siskind}, {Spandre}, {Spinelli}, {Suson}, {Tachibana}, {Thompson}, {Torres},
  {Torresi}, {Troja}, {Valverde}, {van Zyl}, \&
  {Yassine}}]{2020ApJ...892..105A}
{Ajello}, M., {Angioni}, R., {Axelsson}, M., {et~al.} 2020, \apj, 892, 105

\bibitem[{{Akaike}(1974)}]{1974ITAC...19..716A}
{Akaike}, H. 1974, IEEE Transactions on Automatic Control, 19, 716

\bibitem[{{Alarie} {et~al.}(2014){Alarie}, {Bilodeau}, \&
  {Drissen}}]{2014MNRAS.441.2996A}
{Alarie}, A., {Bilodeau}, A., \& {Drissen}, L. 2014, \mnras, 441, 2996

\bibitem[{{Archambault} {et~al.}(2017){Archambault}, {Archer}, {Benbow},
  {Bird}, {Bourbeau}, {Buchovecky}, {Buckley}, {Bugaev}, {Cerruti}, {Connolly},
  {Cui}, {Dwarkadas}, {Errando}, {Falcone}, {Feng}, {Finley}, {Fleischhack},
  {Fortson}, {Furniss}, {Griffin}, {H{\"u}tten}, {Hanna}, {Holder}, {Johnson},
  {Kaaret}, {Kar}, {Kelley-Hoskins}, {Kertzman}, {Kieda}, {Krause}, {Kumar},
  {Lang}, {Maier}, {McArthur}, {McCann}, {Moriarty}, {Mukherjee}, {Nieto},
  {O'Brien}, {Ong}, {Otte}, {Park}, {Pohl}, {Popkow}, {Pueschel}, {Quinn},
  {Ragan}, {Reynolds}, {Richards}, {Roache}, {Sadeh}, {Santander}, {Sembroski},
  {Shahinyan}, {Slane}, {Staszak}, {Telezhinsky}, {Trepanier}, {Tyler},
  {Wakely}, {Weinstein}, {Weisgarber}, {Wilcox}, {Wilhelm}, {Williams}, \&
  {Zitzer}}]{2017ApJ...836...23A}
{Archambault}, S., {Archer}, A., {Benbow}, W., {et~al.} 2017, \apj, 836, 23

\bibitem[{{Astropy Collaboration} {et~al.}(2013){Astropy Collaboration},
  {Robitaille}, {Tollerud}, {Greenfield}, {Droettboom}, {Bray}, {Aldcroft},
  {Davis}, {Ginsburg}, {Price-Whelan}, {Kerzendorf}, {Conley}, {Crighton},
  {Barbary}, {Muna}, {Ferguson}, {Grollier}, {Parikh}, {Nair}, {Unther},
  {Deil}, {Woillez}, {Conseil}, {Kramer}, {Turner}, {Singer}, {Fox}, {Weaver},
  {Zabalza}, {Edwards}, {Azalee Bostroem}, {Burke}, {Casey}, {Crawford},
  {Dencheva}, {Ely}, {Jenness}, {Labrie}, {Lim}, {Pierfederici}, {Pontzen},
  {Ptak}, {Refsdal}, {Servillat}, \& {Streicher}}]{astropy:2013}
{Astropy Collaboration}, {Robitaille}, T.~P., {Tollerud}, E.~J., {et~al.} 2013,
  \aap, 558, A33

\bibitem[{{Atwood} {et~al.}(2009){Atwood}, {Abdo}, {Ackermann}, {Althouse},
  {Anderson}, {Axelsson}, {Baldini}, {Ballet}, {Band}, {Barbiellini}, \&
  et~al.}]{2009ApJ...697.1071A}
{Atwood}, W.~B., {Abdo}, A.~A., {Ackermann}, M., {et~al.} 2009, \apj, 697, 1071

\bibitem[{{Bamba} {et~al.}(2005){Bamba}, {Yamazaki}, {Yoshida}, {Terasawa}, \&
  {Koyama}}]{2005ApJ...621..793B}
{Bamba}, A., {Yamazaki}, R., {Yoshida}, T., {Terasawa}, T., \& {Koyama}, K.
  2005, \apj, 621, 793

\bibitem[{{Bandiera}(1987)}]{1987ApJ...319..885B}
{Bandiera}, R. 1987, \apj, 319, 885

\bibitem[{{Blair} {et~al.}(2007){Blair}, {Ghavamian}, {Long}, {Williams},
  {Borkowski}, {Reynolds}, \& {Sankrit}}]{2007ApJ...662..998B}
{Blair}, W.~P., {Ghavamian}, P., {Long}, K.~S., {et~al.} 2007, \apj, 662, 998

\bibitem[{{Burkey} {et~al.}(2013){Burkey}, {Reynolds}, {Borkowski}, \&
  {Blondin}}]{2013ApJ...764...63B}
{Burkey}, M.~T., {Reynolds}, S.~P., {Borkowski}, K.~J., \& {Blondin}, J.~M.
  2013, \apj, 764, 63

\bibitem[{{Cassam-Chena{\"\i}} {et~al.}(2004){Cassam-Chena{\"\i}},
  {Decourchelle}, {Ballet}, {Hwang}, {Hughes}, {Petre}, \& {et
  al.}}]{2004A&A...414..545C}
{Cassam-Chena{\"\i}}, G., {Decourchelle}, A., {Ballet}, J., {et~al.} 2004,
  \aap, 414, 545

\bibitem[{{Castelletti} {et~al.}(2021){Castelletti}, {Supan}, {Peters}, \&
  {Kassim}}]{2021A&A...653A..62C}
{Castelletti}, G., {Supan}, L., {Peters}, W.~M., \& {Kassim}, N.~E. 2021, \aap,
  653, A62

\bibitem[{{Cherenkov Telescope Array Consortium} {et~al.}(2019){Cherenkov
  Telescope Array Consortium}, {Acharya}, {Agudo}, {Al Samarai}, {Alfaro},
  {Alfaro}, {Alispach}, {Alves Batista}, {Amans}, {Amato}, \&
  et~al.}]{2019scta.book.....C}
{Cherenkov Telescope Array Consortium}, {Acharya}, B.~S., {Agudo}, I., {et~al.}
  2019, {Science with the Cherenkov Telescope Array}

\bibitem[{{Deil} {et~al.}(2017){Deil}, {Zanin}, {Lefaucheur}, {Boisson},
  {Khelifi}, {Terrier}, {Wood}, {Mohrmann}, {Chakraborty}, {Watson},
  {Lopez-Coto}, {Klepser}, {Cerruti}, {Lenain}, {Acero}, {Djannati-Ata{\"\i}},
  {Pita}, {Bosnjak}, {Trichard}, {Vuillaume}, {Donath}, {Consortium}, {King},
  {Jouvin}, {Owen}, {Sipocz}, {Lennarz}, {Voruganti}, {Spir-Jacob}, {Ruiz}, \&
  {Arribas}}]{gammapy:2017}
{Deil}, C., {Zanin}, R., {Lefaucheur}, J., {et~al.} 2017, in International
  Cosmic Ray Conference, Vol. 301, 35th International Cosmic Ray Conference
  (ICRC2017), 766

\bibitem[{{DeLaney} {et~al.}(2002){DeLaney}, {Koralesky}, {Rudnick}, \&
  {Dickel}}]{2002ApJ...580..914D}
{DeLaney}, T., {Koralesky}, B., {Rudnick}, L., \& {Dickel}, J.~R. 2002, \apj,
  580, 914

\bibitem[{{Diesing} \& {Caprioli}(2021)}]{2021ApJ...922....1D}
{Diesing}, R. \& {Caprioli}, D. 2021, \apj, 922, 1

\bibitem[{{Fesen} {et~al.}(2018){Fesen}, {Weil}, {Cisneros}, {Blair}, \&
  {Raymond}}]{2018MNRAS.481.1786F}
{Fesen}, R.~A., {Weil}, K.~E., {Cisneros}, I.~A., {Blair}, W.~P., \& {Raymond},
  J.~C. 2018, \mnras, 481, 1786

\bibitem[{{H.~E.~S.~S. Collaboration} {et~al.}(2018){H.~E.~S.~S.
  Collaboration}, {Abdalla}, {Abramowski}, {Aharonian}, {Ait Benkhali},
  {Ang{\"u}ner}, {Arakawa}, {Arrieta}, {Aubert}, {Backes}, \&
  et~al.}]{2018A&A...612A...1H}
{H.~E.~S.~S. Collaboration}, {Abdalla}, H., {Abramowski}, A., {et~al.} 2018,
  \aap, 612, A1

\bibitem[{{Hayato} {et~al.}(2010){Hayato}, {Yamaguchi}, {Tamagawa}, {Katsuda},
  {Hwang}, {Hughes}, {Ozawa}, {Bamba}, {Kinugasa}, {Terada}, {Furuzawa},
  {Kunieda}, \& {Makishima}}]{2010ApJ...725..894H}
{Hayato}, A., {Yamaguchi}, H., {Tamagawa}, T., {et~al.} 2010, \apj, 725, 894

\bibitem[{{Humensky} \& {VERITAS Collaboration}(2015)}]{2015ICRC...34..875H}
{Humensky}, B. \& {VERITAS Collaboration}. 2015, in International Cosmic Ray
  Conference, Vol.~34, 34th International Cosmic Ray Conference (ICRC2015), 875

\bibitem[{{Kafexhiu} {et~al.}(2014){Kafexhiu}, {Aharonian}, {Taylor}, \&
  {Vila}}]{2014PhRvD..90l3014K}
{Kafexhiu}, E., {Aharonian}, F., {Taylor}, A.~M., \& {Vila}, G.~S. 2014, \prd,
  90, 123014

\bibitem[{{Katsuda} {et~al.}(2015){Katsuda}, {Mori}, {Maeda}, {Tanaka},
  {Koyama}, {Tsunemi}, {Nakajima}, {Maeda}, {Ozaki}, \&
  {Petre}}]{2015ApJ...808...49K}
{Katsuda}, S., {Mori}, K., {Maeda}, K., {et~al.} 2015, \apj, 808, 49

\bibitem[{{Katsuda} {et~al.}(2008){Katsuda}, {Tsunemi}, {Uchida}, \&
  {Kimura}}]{2008ApJ...689..225K}
{Katsuda}, S., {Tsunemi}, H., {Uchida}, H., \& {Kimura}, M. 2008, \apj, 689,
  225

\bibitem[{{Nagayoshi} {et~al.}(2021){Nagayoshi}, {Bamba}, {Katsuda}, \&
  {Terada}}]{2021PASJ..tmp...10N}
{Nagayoshi}, T., {Bamba}, A., {Katsuda}, S., \& {Terada}, Y. 2021, \pasj

\bibitem[{{Nigro} {et~al.}(2019){Nigro}, {Deil}, {Zanin}, {Hassan}, {King},
  {Ruiz}, {Saha}, {Terrier}, {Br{\"u}gge}, {N{\"o}the}, {Bird}, {Lin},
  {Aleksi{\'c}}, {Boisson}, {Contreras}, {Donath}, {Jouvin}, {Kelley-Hoskins},
  {Khelifi}, {Kosack}, {Rico}, \& {Sinha}}]{gammapy:2019}
{Nigro}, C., {Deil}, C., {Zanin}, R., {et~al.} 2019, \aap, 625, A10

\bibitem[{{Parizot} {et~al.}(2006){Parizot}, {Marcowith}, {Ballet}, \&
  {Gallant}}]{2006A&A...453..387P}
{Parizot}, E., {Marcowith}, A., {Ballet}, J., \& {Gallant}, Y.~A. 2006, \aap,
  453, 387

\bibitem[{{Porter} {et~al.}(2006){Porter}, {Moskalenko}, \&
  {Strong}}]{2006ApJ...648L..29P}
{Porter}, T.~A., {Moskalenko}, I.~V., \& {Strong}, A.~W. 2006, \apjl, 648, L29

\bibitem[{{Price-Whelan} {et~al.}(2018){Price-Whelan}, {Sip{\H{o}}cz},
  {G{\"u}nther}, {Lim}, {Crawford}, {Conseil}, {Shupe}, {Craig}, {Dencheva},
  {Ginsburg}, {VanderPlas}, {Bradley}, {P{\'e}rez-Su{\'a}rez}, {de Val-Borro},
  {Paper Contributors}, {Aldcroft}, {Cruz}, {Robitaille}, {Tollerud},
  {Coordination Committee}, {Ardelean}, {Babej}, {Bach}, {Bachetti}, {Bakanov},
  {Bamford}, {Barentsen}, {Barmby}, {Baumbach}, {Berry}, {Biscani}, {Boquien},
  {Bostroem}, {Bouma}, {Brammer}, {Bray}, {Breytenbach}, {Buddelmeijer},
  {Burke}, {Calderone}, {Cano Rodr{\'\i}guez}, {Cara}, {Cardoso}, {Cheedella},
  {Copin}, {Corrales}, {Crichton}, {D{\textquoteright}Avella}, {Deil},
  {Depagne}, {Dietrich}, {Donath}, {Droettboom}, {Earl}, {Erben}, {Fabbro},
  {Ferreira}, {Finethy}, {Fox}, {Garrison}, {Gibbons}, {Goldstein}, {Gommers},
  {Greco}, {Greenfield}, {Groener}, {Grollier}, {Hagen}, {Hirst}, {Homeier},
  {Horton}, {Hosseinzadeh}, {Hu}, {Hunkeler}, {Ivezi{\'c}}, {Jain}, {Jenness},
  {Kanarek}, {Kendrew}, {Kern}, {Kerzendorf}, {Khvalko}, {King}, {Kirkby},
  {Kulkarni}, {Kumar}, {Lee}, {Lenz}, {Littlefair}, {Ma}, {Macleod},
  {Mastropietro}, {McCully}, {Montagnac}, {Morris}, {Mueller}, {Mumford},
  {Muna}, {Murphy}, {Nelson}, {Nguyen}, {Ninan}, {N{\"o}the}, {Ogaz}, {Oh},
  {Parejko}, {Parley}, {Pascual}, {Patil}, {Patil}, {Plunkett}, {Prochaska},
  {Rastogi}, {Reddy Janga}, {Sabater}, {Sakurikar}, {Seifert}, {Sherbert},
  {Sherwood-Taylor}, {Shih}, {Sick}, {Silbiger}, {Singanamalla}, {Singer},
  {Sladen}, {Sooley}, {Sornarajah}, {Streicher}, {Teuben}, {Thomas},
  {Tremblay}, {Turner}, {Terr{\'o}n}, {van Kerkwijk}, {de la Vega}, {Watkins},
  {Weaver}, {Whitmore}, {Woillez}, {Zabalza}, \& {Contributors}}]{astropy:2018}
{Price-Whelan}, A.~M., {Sip{\H{o}}cz}, B.~M., {G{\"u}nther}, H.~M., {et~al.}
  2018, \aj, 156, 123

\bibitem[{{Prokhorov} {et~al.}(2021){Prokhorov}, {Vink}, {Simoni}, {Komin},
  {Funk}, {Malyshev}, {Mohrmann}, {Ohm}, {P{\"u}hlhofer}, \&
  {V{\"o}lk}}]{2021icrc}
{Prokhorov}, D., {Vink}, J., {Simoni}, R., {et~al.} 2021, Proceedings of the
  37th International Cosmic Ray Conference (ICRC 2021), arXiv:2107.11582

\bibitem[{{Ranasinghe} \& {Leahy}(2018)}]{2018AJ....155..204R}
{Ranasinghe}, S. \& {Leahy}, D.~A. 2018, \aj, 155, 204

\bibitem[{{Rettig} \& {Pohl}(2012)}]{2012A&A...545A..47R}
{Rettig}, R. \& {Pohl}, M. 2012, \aap, 545, A47

\bibitem[{{Reynolds} {et~al.}(2007){Reynolds}, {Borkowski}, {Hwang}, {Hughes},
  {Badenes}, {Laming}, \& {Blondin}}]{2007ApJ...668L.135R}
{Reynolds}, S.~P., {Borkowski}, K.~J., {Hwang}, U., {et~al.} 2007, \apjl, 668,
  L135

\bibitem[{{Reynoso} \& {Goss}(1999)}]{1999AJ....118..926R}
{Reynoso}, E.~M. \& {Goss}, W.~M. 1999, \aj, 118, 926

\bibitem[{{Sankrit} {et~al.}(2005){Sankrit}, {Blair}, {Delaney}, {Rudnick},
  {Harrus}, \& {Ennis}}]{2005AdSpR..35.1027S}
{Sankrit}, R., {Blair}, W.~P., {Delaney}, T., {et~al.} 2005, Advances in Space
  Research, 35, 1027

\bibitem[{{Sankrit} {et~al.}(2016){Sankrit}, {Raymond}, {Blair}, {Long},
  {Williams}, {Borkowski}, {Patnaude}, \& {Reynolds}}]{2016ApJ...817...36S}
{Sankrit}, R., {Raymond}, J.~C., {Blair}, W.~P., {et~al.} 2016, \apj, 817, 36

\bibitem[{{Tsuji} {et~al.}(2021){Tsuji}, {Uchiyama}, {Khangulyan}, \&
  {Aharonian}}]{2021ApJ...907..117T}
{Tsuji}, N., {Uchiyama}, Y., {Khangulyan}, D., \& {Aharonian}, F. 2021, \apj,
  907, 117

\bibitem[{{Vink}(2008)}]{2008ApJ...689..231V}
{Vink}, J. 2008, \apj, 689, 231

\bibitem[{{Vink}(2017)}]{2017hsn..book..139V}
{Vink}, J. 2017, {Supernova 1604, Kepler's Supernova, and its Remnant}, 139

\bibitem[{{Wood} {et~al.}(2017){Wood}, {Caputo}, {Charles}, {Di Mauro},
  {Magill}, {Perkins}, \& {Fermi-LAT Collaboration}}]{2017ICRC...35..824W}
{Wood}, M., {Caputo}, R., {Charles}, E., {et~al.} 2017, in International Cosmic
  Ray Conference, Vol. 301, 35th International Cosmic Ray Conference
  (ICRC2017), 824

\bibitem[{{Xiang} \& {Jiang}(2021)}]{2021ApJ...908...22X}
{Xiang}, Y. \& {Jiang}, Z. 2021, \apj, 908, 22

\bibitem[{{Yoshiike} {et~al.}(2013){Yoshiike}, {Fukuda}, {Sano}, {Ohama},
  {Moribe}, {Torii}, {Hayakawa}, {Okuda}, {Yamamoto}, {Tajima}, {Mizuno},
  {Nishimura}, {Kimura}, {Maezawa}, {Onishi}, {Mizuno}, {Ogawa}, {Giuliani},
  {Koo}, \& {Fukui}}]{2013ApJ...768..179Y}
{Yoshiike}, S., {Fukuda}, T., {Sano}, H., {et~al.} 2013, \apj, 768, 179

\bibitem[{{Yu} {et~al.}(2019){Yu}, {Chen}, {Jiang}, \&
  {Zijlstra}}]{2019MNRAS.488.3129Y}
{Yu}, B., {Chen}, B.~Q., {Jiang}, B.~W., \& {Zijlstra}, A. 2019, \mnras, 488,
  3129

\bibitem[{{Zabalza}(2015)}]{2015ICRC...34..922Z}
{Zabalza}, V. 2015, in International Cosmic Ray Conference, Vol.~34, 34th
  International Cosmic Ray Conference (ICRC2015), 922

\bibitem[{{Zirakashvili} \& {Aharonian}(2007)}]{2007A&A...465..695Z}
{Zirakashvili}, V.~N. \& {Aharonian}, F. 2007, \aap, 465, 695

\end{thebibliography}
\end{document}